# Error Rates of Capacity-Achieving Codes Are Convex

Sergey Loyka

SITE
University of Ottawa
Ottawa, K1N 6N5, Canada
e-mail: sergey.loyka@ieee.org

Francois Gagnon

Department of Electrical Engineering
Ecole de Technologie Superieure
Montreal, H3C 1K3, Canada
e-mail: francois.gagnon@etsmtl.ca

Victoria Kostina

Department of Electrical Engineering
Princeton University
Princeton, NJ, 08544, USA
e-mail: vkostina@princeton.edu

*Abstract*— Motivated by a wide-spread use of convex optimization techniques, convexity properties of bit error rate of the maximum likelihood detector operating in the AWGN channel are studied for arbitrary constellations and bit mappings, which also includes coding under maximum-likelihood decoding. Under this generic setting, the pairwise probability of error and bit error rate are shown to be convex functions of the SNR and noise power in the high SNR/low noise regime with explicitly-determined boundary. Any code, including capacity-achieving ones, whose decision regions include the hardened noise spheres (from the noise sphere hardening argument in the channel coding theorem) satisfies this high SNR requirement and thus has convex error rates in both SNR and noise power. We conjecture that *all* capacity-achieving codes have convex error rates.

I. INTRODUCTION

Optimization problems of various kinds simplify significantly if the goal and constraint functions involved are convex. Indeed, a convex optimization problem has a unique global solution, which can be found either analytically or, with a reasonable effort, by several efficient numerical methods; its numerical complexity grows only moderately with the problem dimensionality and required accuracy; convergence rates and required step size can be estimated in advance; there are powerful analytical tools that can be used to attack a problem and that provide insights into such problems even if solutions, either analytical or numerical, are not found yet [1][2]. In a sense, convex problems are as easy as linear ones. Contrary to this, generic nonlinear optimization problems not only possess none of these features, but also are not solvable numerically, i.e. their complexity grows prohibitively fast with problem dimensionality and required accuracy [2]. Thus, there is a great advantage in formulating or at least in approximating an optimization problem as a convex one.

In the world of digital communications, one of the major performance measures is either symbol error rate (SER) or bit error rate (BER). Consequently, when an optimization of a communication system is performed, either SER or BER often appears as the goal or constraint function. Examples include optimum power/rate allocation in spatial multiplexing systems (BLAST) [3], optimum power/time sharing for a transmitter and a jammer [4], rate allocation or precoding in multicarrier (OFDM) systems [5], optimum equalization [6], optimum multiuser detection [7], and joint Tx-Rx beamforming (precoding-decoding) in MIMO systems [8]. Symbol and bit error rates of the maximum likelihood (ML) detector have been extensively studied and a large number of exact or approximate analytical results are available for various modulation formats, for both non-fading and fading AWGN channels [9]. On the other hand, convexity properties of error rates are not understood so well, especially for constellations of complicated geometry, large dimensionality or when coding is used. Results in this area are scarce. Many known closed-form error rate expressions can be verified by differentiation to be convex, but this approach does not lead anywhere in general. Convexity properties for binary modulations have been studied in-depth in [4], including applications to transmitter and jammer optimizations, and were later extended to arbitrary multidimensional constellations in [10][11] in terms of the SER under maximum-likelihood detection.

Unfortunately, convexity of SER does not say anything in general about convexity of the BER, since the latter depends on pairwise probabilities of error (PEP) and not on the SER [12]. Since the BER is an important performance indicator and thus appears as an objective in many optimization problems, we study its convexity in the present paper using a generic geometrical framework developed in [10][11]. Our setting is generic enough so that the results apply to constellations of arbitrary order, shape and dimensionality, which also includes coding under maximum likelihood decoding.

First, we establish convexity properties of the PEP as a function of SNR: it is convex at the high SNR regime for any constellation/coding. Its low-SNR behavior depends on constellation dimensionality: it is concave in dimensions 1 and 2 with an odd number of inflection points at intermediate SNR, and it is convex in higher dimensions with an even number of inflection points. Based on this, convexity of the BER at high SNR is established for arbitrary constellation, bit mapping and coding. Thus, this property is a consequence of Gaussian noise density and maximum likelihood detection rather than particular constellation, bit mapping or coding technique. We also show that the BER is a convex function of the noise power in the small noise/high SNR mode.

While the convexity of PEP and BER has been established at high SNR, the question remains: how relevant this high SNR region is, i.e. does it correspond to realistic/practical SNR values? This has significant impact on the result's importance and its utility when solving practically-relevant optimization problems. In this paper, we provide a positive answer: the high SNR is almost the same as that required by the channel coding theorem so that any code, including capacity-achieving ones, whose decision regions include the hardened noise spheres

(from the sphere packing/hardening arguments in the channel coding theorem [9][13]) are in this range. In other words, our boundary of the high SNR regime is closely matched to that in the channel coding theorem so that arbitrary low probability of error implies its convexity. Any practical code whose decision regions include the hardened noise spheres has also convex SER, PEP and BER. This opens up an opportunity to apply numerous and powerful tools of convex optimization to design of systems with such codes on a rigorous basis.

## II. SYSTEM MODEL

The standard baseband discrete-time system model with an AWGN channel, which includes matched filtering and sampling, is

$$\mathbf{r} = \mathbf{s} + \boldsymbol{\xi} \quad (1)$$

where $\mathbf{s}$ and $\mathbf{r}$ are $n$-dimensional vectors representing the Tx and Rx symbols respectively, $\mathbf{s} \in \{\mathbf{s}_1, \mathbf{s}_2, ..., \mathbf{s}_M\}$, a set of $M$ constellation points, $\boldsymbol{\xi}$ is the additive white Gaussian noise (AWGN), $\boldsymbol{\xi} \sim \mathcal{N}(\mathbf{0}, \sigma_0^2 \mathbf{I})$, whose probability density function (PDF) is

$$p_\xi(\mathbf{x}) = \left(2\pi\sigma_0^2\right)^{-n/2} e^{-|\mathbf{x}|^2/2\sigma_0^2} \quad (2)$$

where $\sigma_0^2$ is the noise variance per dimension, and $n$ is the constellation dimensionality; lower case bold letters denote vectors, bold capitals denote matrices, $x_i$ denotes i-th component of $\mathbf{x}$, $|\mathbf{x}|$ denotes $L_2$ norm of $\mathbf{x}$, $|\mathbf{x}| = \sqrt{\mathbf{x}^T \mathbf{x}}$, where the superscript $T$ denotes transpose, $\mathbf{x}_i$ denotes i-th vector. The average (over the constellation points) SNR is defined as $\gamma = 1/\sigma_0^2$, which implies the appropriate normalization, $\frac{1}{M}\sum_{i=1}^{M} |\mathbf{s}_i|^2 = 1$.

Consider the maximum likelihood detector, which is equivalent to the minimum distance one in the AWGN channel, $\hat{\mathbf{s}} = \arg\min_{\mathbf{s}_i} |\mathbf{r} - \mathbf{s}_i|$. The probability of symbol error $P_{ei}$ given that $\mathbf{s} = \mathbf{s}_i$ was transmitted is $P_{ei} = \Pr[\hat{\mathbf{s}} \neq \mathbf{s}_i | \mathbf{s} = \mathbf{s}_i] = 1 - P_{ci}$, where $P_{ci}$ is the probability of correct decision. The SER averaged over all constellation points is $P_e = \sum_{i=1}^{M} P_{ei} \Pr[\mathbf{s} = \mathbf{s}_i] = 1 - P_c$. $P_{ei}$ can be expressed as

$$P_{ei} = 1 - \int_{\Omega_i} p_\xi(\mathbf{x}) d\mathbf{x} \quad (3)$$

where $\Omega_i$ is the decision region (Voronoi region), and $\mathbf{s}_i$ corresponds to $\mathbf{x} = 0$, i.e. the origin is shifted for convenience to the constellation point $\mathbf{s}_i$. $\Omega_i$ can be expressed as a convex polyhedron [1],

$$\Omega_i = \{\mathbf{x} | \mathbf{A}\mathbf{x} \leq \mathbf{b}\}, \quad \mathbf{a}_j^T = \frac{(\mathbf{s}_j - \mathbf{s}_i)}{|\mathbf{s}_j - \mathbf{s}_i|}, \quad b_j = \frac{1}{2}|\mathbf{s}_j - \mathbf{s}_i| \quad (4)$$

where $\mathbf{a}_j^T$ denotes j-th row of $\mathbf{A}$, and the inequality in (4) is applied component-wise. Clearly, $P_{ei}$ and $P_{ci}$ possess the opposite convexity properties.

Another important performance indicator is the pairwise probability of error (PEP) i.e. a probability $\Pr\{\mathbf{s}_i \to \mathbf{s}_j\} = \Pr[\hat{\mathbf{s}} = \mathbf{s}_j | \mathbf{s} = \mathbf{s}_i]$ to decide in favor of $\mathbf{s}_j$ given that $\mathbf{s}_i$, $i \neq j$, was transmitted, which can be expressed as

$$\Pr\{\mathbf{s}_i \to \mathbf{s}_j\} = \int_{\Omega_j} p_\xi(\mathbf{x}) d\mathbf{x} \quad (5)$$

where $\Omega_j$ is the decision region for $\mathbf{s}_j$ when the reference frame is centered at $\mathbf{s}_i$. The SER can now be expressed as

$$P_{ei} = \sum_{j \neq i} \Pr\{\mathbf{s}_i \to \mathbf{s}_j\} \quad (6)$$

and the BER can be expressed as a positive linear combination of PEPs [12]

$$\text{BER} = \sum_{i=1}^{M} \sum_{j \neq i} \frac{h_{ij}}{\log_2 M} \Pr\{\mathbf{s} = \mathbf{s}_i\} \Pr\{\mathbf{s}_i \to \mathbf{s}_j\} \quad (7)$$

where $h_{ij}$ is the Hamming distance between binary sequences representing $\mathbf{s}_i$ and $\mathbf{s}_j$.

Note that the model and error rate expressions we are using are generic enough to apply to arbitrary constellations, which also includes coding under maximum-likelihood decoding (codewords are considered as points of an extended constellation). We now proceed to establish convexity properties of error rates in this generic setting.

## III. CONVEXITY OF SYMBOL ERROR RATES

Convexity properties of symbol error rates of the ML detector in the SNR and noise power have been established in [10][11] for arbitrary constellation/coding (under ML decoding) and are summarized below for completeness and comparison purpose.

**Theorem 1 (Theorem 1 and 2 in [10])**: The SER $P_e$ is a convex function of the SNR $\gamma$ for any constellation/coding (under ML decoding) if $n \leq 2$,

$$d^2 P_e / d\gamma^2 = P''_{e|\gamma} > 0 \quad (8)$$

For $n > 2$, the following convexity properties hold:
- $P_{ei}$ is convex in the large SNR mode,

$$\gamma \geq \left(n + \sqrt{2n}\right) / d_{\min,i}^2 \quad (9)$$

where $d_{\min,i}$ is the minimum distance from $\mathbf{s}_i$ to its decision region boundary,

- $P_{ei}$ is concave in the low SNR mode,

$$\gamma \leq \left(n - \sqrt{2n}\right) / d_{\max,i}^2 \quad (10)$$

where $d_{\max,i}$ is the maximum distance from $\mathbf{s}_i$ to its decision region boundary,

- there are an odd number of inflection points, $P''_{ci|\gamma} = P''_{ei|\gamma} = 0$, in the intermediate SNR mode,

$$\left(n - \sqrt{2n}\right) / d_{\max,i}^2 \leq \gamma \leq \left(n + \sqrt{2n}\right) / d_{\min,i}^2 \quad (11)$$

- the SER $P_e$ is convex at high SNR,

$$\gamma \geq \left(n + \sqrt{2n}\right) / d_{\min}^2 \quad (12)$$

where $d_{\min} = \min_i \{d_{\min,i}\}$ is the minimum distance to decision region boundary in the constellation.

While Theorem 1 does not conclude that the SER $P_e$ is concave at low SNR, examples of constellations when this is indeed so can be found in [14].

**Theorem 2 (Theorem 4 in [10]):** Symbol error rates have the following convexity properties in the noise power $\sigma_0^2$, for any constellation/coding,

- $P_{ei}$ is concave in the large noise (low SNR) mode,
$$\left(n+2-\sqrt{2(n+2)}\right)\sigma_0^2 \geq d_{\max,i}^2 \quad (13)$$

- $P_{ei}$ is convex in the small noise (high SNR) mode,
$$\left(n+2+\sqrt{2(n+2)}\right)\sigma_0^2 \leq d_{\min,i}^2 \quad (14)$$

- there are an odd number of inflection points for intermediate noise power,
$$d_{\min,i}^2 \left(n+2+\sqrt{2(n+2)}\right)^{-1} \leq \sigma_0^2 \leq d_{\max,i}^2 \left(n+2-\sqrt{2(n+2)}\right)^{-1} \quad (15)$$

- the SER $P_e$ is convex in the small noise/high SNR mode,
$$\left(n+2+\sqrt{2(n+2)}\right)\sigma_0^2 \leq d_{\min}^2 \quad (16)$$

While the convexity properties above are important for many optimization problems, they do not lend any conclusions about convexity of the BER, since the latter is not directly related to $P_e$ or $P_{ei}$ in general. While, in some cases, the BER can be expressed as linear combination of $P_{ei}$, there are positive and negative terms so that no conclusion about convexity can be made in this case either. On the other hand, the BER can be expressed as a positive linear combination of pairwise probabilities of error so that the convexity of the latter implies the convexity of the former. Thus, we study below the convexity property of the PEP, from which the convexity property of the BER will follow.

## IV. CONVEXITY OF PEP AND BER

In many cases, it is a pairwise error probability that is a key point in the analysis (e.g. for constructing a union bound and other performance metrics). Furthermore, it is also a basic building block for the BER in (7), so that we establish its convexity property first.

**Theorem 3**:

a) For any constellation/code, the pairwise error probability $\Pr\{\mathbf{s}_i \to \mathbf{s}_j\}$ is a convex function of the SNR at the high SNR/low noise region, $\gamma \geq (n+\sqrt{2n})/d_{\min,i}^2$ or, equivalently, $d_{\min,i}^2 \geq (n+\sqrt{2n})\sigma_0^2$;

b) for $n=1,2$, it is concave at the low SNR region, $\gamma \leq (n+\sqrt{2n})/(d_{ij}+d_{\max,j})^2$, where $d_{ij}=|\mathbf{s}_i-\mathbf{s}_j|$ is the distance between $\mathbf{s}_i$ and $\mathbf{s}_j$, and there is an odd number of inflection points, $\Pr\{\mathbf{s}_i \to \mathbf{s}_j\}'' = 0$, in the intermediate SNR mode,
$$(n+\sqrt{2n})/(d_{ij}+d_{\max,j})^2 \leq \gamma \leq (n+\sqrt{2n})/d_{\min,i}^2 \quad (17)$$

c) for $n > 2$, the PEP is convex at the low SNR region, $\gamma \leq (n-\sqrt{2n})/(d_{ij}+d_{\max,j})^2$, and there is an even number of inflection points in-between,
$$(n-\sqrt{2n})/(d_{ij}+d_{\max,j})^2 \leq \gamma \leq (n+\sqrt{2n})/d_{\min,i}^2$$

**Proof:** See Appendix.

We note that Theorem 3(a) is stronger than Theorem 1 at the high SNR region since the latter follows from the former but not the other way around (as the other SNR ranges in Theorem 3 above indicate). Unlike the SER, the pairwise error probability can be concave at low SNR even for $n=1,2$.

Since Theorem 3 holds for any constellation and bit mapping, it follows that the convexity property of the PEP at high SNR is a consequence of Gaussian noise density rather than particular modulation/coding used, where the latter determines only the high SNR threshold via the dimensionality and minimum distance.

We are now in a position to establish the convexity of bit error rate.

**Theorem 4**: The BER is a convex function of the SNR, for any constellation and bit mapping, which also includes coding under maximum-likelihood decoding, at the high SNR/small noise regime,
$$d_{\min}^2 \geq (n+\sqrt{2n})\sigma_0^2, \quad (18)$$

where $d_{\min} = \min_i\{d_{\min,i}\}$ is the minimum distance to the boundary in the constellation, and the SNR $\gamma = 1/\sigma_0^2$.

**Proof:** Using the relationship between the BER and the pairwise error probabilities in (7) and observing that a positive linear combination of convex functions is convex. Q.E.D.

Thus, the condition in (18) guarantees the convexity of all PEP, BER and SER. In some cases (Gray encoding and when nearest neighbor errors dominate), the BER is approximated as $SER/\log_2 M$, so that it inherits the same convexity properties as in Theorems 1 and 2 above.

We remark that the lower bound in (18) has an interesting interpretation: $n\sigma_0^2$ is the mean of $|\xi|^2$ and $\sqrt{2n}\sigma_0^2$ is its standard deviation, so that (18) requires that $d_{\min}^2$ be larger than the average noise power by at least its standard deviation, which is intuitively what is required for low probability of error. Thus, the condition in (18) should be satisfied when probability of error is small. The next section makes this statement more precise.

## V. HOW HIGH IS THE HIGH SNR?

We now proceed to establish practical relevance of the high-SNR regime in (18) based on the channel coding theorem. Recall that the sphere hardening argument (from the channel coding theorem) states that the noise vector $\xi$ is contained within the sphere of radius $\sqrt{n(\sigma_0^2+\varepsilon)}$ with high probability (approaching 1 as $n \to \infty$) [13][9], where $\varepsilon > 0$ is a fixed, arbitrary small number, so that the decision regions should have minimum distance to the boundary
$$d_{\min} \geq \sqrt{n(\sigma_0^2+\varepsilon)}, \quad (19)$$

i.e. to enclose the hardened noise sphere of radius $\sqrt{n(\sigma_0^2+\varepsilon)}$, to provide arbitrary low probability of error as $n \to \infty$. For any code satisfying this requirement, it follows that
$$d_{\min}^2 \geq n(\sigma_0^2+\varepsilon) > (n+\sqrt{2n})\sigma_0^2, \quad (20)$$

for sufficiently large $n$ and $\forall \varepsilon > 0$. Thus, for any code whose decision regions enclose the hardened noise spheres, the condition of Theorem 4 is satisfied and therefore the error rates (SER, PEP, BER) of such capacity-approaching codes are all convex.

On the other hand, for any code whose decisions regions are enclosed by the spheres of radius $\sqrt{n+\sqrt{2n}}\sigma_0$ (i.e. completely violate (18)), the symbol error rates are lower bounded as

$$P_{ei} \geq \Pr\left\{\frac{|\xi|-n}{\sqrt{2n}} > 1\right\} \approx Q(1) \approx 0.16 > 0, \quad (21)$$

where $Q(x) = \frac{1}{\sqrt{2\pi}}\int_x^\infty e^{-t^2/2} dt$ is the Q-function, so that low probability of error is not achievable. Based on these two arguments, we conjecture the following.

**Conjecture:** Consider a capacity-achieving code designed for $\text{SNR} = \gamma_0$. Error rates of *any* such code are convex for $\text{SNR} \geq \gamma_0$, i.e. when it provides an arbitrary low probability of error.

This conjecture is stronger that our convexity statement above since the latter requires the decision regions to include the hardened noise spheres, which is only a sufficient condition for arbitrarily low probability of error. To the best of our knowledge, its necessity has not been established, so that it's possible that a capacity-achieving code violates the condition in (20) (but it has to respect $d_{\max} > \sqrt{n+\sqrt{2n}}\sigma_0$ to avoid (21)). The conjecture effectively states that, if present, such a violation is minor in nature and does not affect the convexity property.

As an application of this result, we note that power/time sharing cannot reduce error rates of any code for which (18) holds. This complements the well-known result that power/time sharing cannot increase the capacity.

In summary, any code respecting the noise sphere hardening and hence having low probability of error will also have convex error rates (SER, PEP and BER). This is a way convexity intimately enters into the channel coding theorem.

VI. CONVEXITY OF THE PEP AND BER IN NOISE POWER

In a jammer optimization problem, it is convexity properties in noise power that are important [4]. Motivated by this fact, we study below convexity of the PEP and BER in the noise power.

**Theorem 5:** The PEP $\Pr\{\mathbf{s}_i \to \mathbf{s}_j\}$ is a convex function of the noise power $\sigma_0^2$, for any $n$, in the small noise/high SNR mode,

$$\left(n+2+\sqrt{2(n+2)}\right)\sigma_0^2 \leq d_{\min,i}^2 \quad (22)$$

and in the large noise/low SNR mode,

$$\left(n+2-\sqrt{2(n+2)}\right)\sigma_0^2 \geq (d_{ij}+d_{\max,j})^2 \quad (23)$$

and has an even number of inflection points in-between.

**Proof:** See Appendix.

Based on this Theorem, the following convexity property of the BER is established.

**Corollary 5.1**: For any constellation and bit mapping, which also includes coding under ML decoding, the BER is a convex function of the noise power in the small noise/high SNR mode:

$$\left(n+2+\sqrt{2(n+2)}\right)\sigma_0^2 \leq d_{\min}^2 \quad (24)$$

where specifics of the constellation/code determine only the high-SNR boundary in (24).

For any code respecting the sphere hardening argument,

$$d_{\min}^2 \geq n(\sigma_0^2 + \varepsilon) > \left(n+2+\sqrt{2(n+2)}\right)\sigma_0^2, \quad (25)$$

for sufficiently large $n$, so that the BER is a convex function of the noise power. For such codes, power/time sharing does not help to decrease the BER, but it is always helpful for a jammer whose objective is to increase the BER. A jammer transmission strategy to maximize the SER via a time/power sharing has been presented in [10][11] and, with some modifications, it can also be used to maximize the BER, following the convexity result in Corollary 5.1.

## VIII. APPENDIX

**Proof of Theorem 3:** The pairwise probability of error $P_{ij} = \Pr\{\mathbf{s}_i \to \mathbf{s}_j\}$ can be presented as

$$P_{ij} = \int_{\Omega_j} p_\xi(\mathbf{x}) d\mathbf{x} \tag{26}$$

where $\Omega_j$ is the decision region for $\mathbf{s}_j$ when the reference frame is centered at $\mathbf{s}_i$. Its second derivative in the SNR is

$$P_{ij}'' = \int_{\Omega_j} \frac{d^2 p_\xi(\mathbf{x})}{d\gamma^2} d\mathbf{x} \tag{27}$$

where the derivative is

$$\frac{d^2 p_\xi(\mathbf{x})}{d\gamma^2} = \frac{1}{4}\left(\frac{\gamma}{2\pi}\right)^{n/2} e^{-\gamma|\mathbf{x}|^2/2} f\left(|\mathbf{x}|^2\right) \tag{28}$$

and $f(t) = (t - \alpha_1/\gamma)(t - \alpha_2/\gamma)$, $\alpha_1 = n + \sqrt{2n} > 0$, $\alpha_2 = n - \sqrt{2n} < \alpha_1$. Consider three different cases.

(i) If $d_{\min,i}^2 \geq \alpha_1/\gamma$, where $d_{\min,i} = \min_j(b_j)$ is the minimum distance from the origin to the boundary of $\Omega_i$, then $f(|\mathbf{x}|^2) \geq 0 \ \forall \mathbf{x} \in \Omega_j$ so that the integral in (27) is clearly positive since the integrand is non-negative everywhere in the integration region and positive in some parts of it. Fig. 1 illustrates this case. This is a high SNR mode since $\gamma \geq \alpha_1/d_{\min,i}^2$.

(ii) If $(d_{ij} + d_{\max,j})^2 \leq \alpha_1/\gamma$ and $n = 1, 2$, where $d_{\max,j}$ is the maximum distance from the center of $\Omega_j$ to its boundary, then $f(|\mathbf{x}|^2) \leq 0 \ \forall \mathbf{x} \in \Omega_j$ so that the integral in (27) is clearly negative and the result follows. Fig. 2 illustrates this case. This is a low-SNR mode since $\gamma \leq \alpha_1/(d_{ij} + d_{\max,j})^2$. An odd number of inflection points in Theorem 3(b) follows from the continuity argument ($P_{ij}''$ is a continuous function of the SNR).

(iii) Part (c) follows from the same argument as in (ii). Q.E.D.

**Proof of Theorem 5:** follows the same geometric technique as for Theorem 3. 2nd derivative of the PEP in the noise power $P_N = \sigma_0^2$ can be expressed as

$$\frac{d^2 P_{ij}}{dP_N^2} = \int_{\Omega_j} \frac{d^2 p_\xi(\mathbf{x})}{P_N^2} d\mathbf{x} \tag{29}$$

where

$$\frac{d^2 p_\xi(\mathbf{x})}{dP_N^2} = \frac{1}{4P_N^4}\left(\frac{1}{2\pi P_N}\right)^{\frac{n}{2}} e^{-\frac{|\mathbf{x}|^2}{2P_N}} f^*\left(|\mathbf{x}|^2\right)$$

$$f^*(t) = (t - \beta_1 P_N)(t - \beta_2 P_N), \tag{30}$$

$$\beta_1 = n + 2 + \sqrt{2(n+2)}, \ \beta_2 = n + 2 - \sqrt{2(n+2)}$$

and $\beta_1 > \beta_2 > 0$. Since $f^*(t)$ has the same structure as $f(t)$ in (28), the proof follows the same steps. In particular, if $d_{\min,i}^2 \geq \beta_1 P_N$, then $d^2 p_\xi / dP_N^2 > 0 \ \forall \mathbf{x} \in \Omega_j$ so that the integral in (29) is clearly positive. The other case is proved in a similar way. Q.E.D.

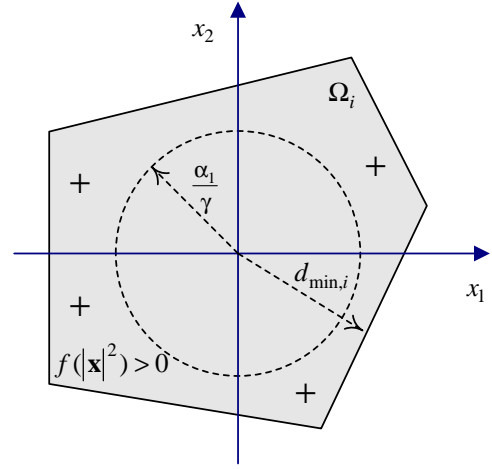

**Fig. 1.** Two-dimensional illustration of the problem geometry for Case 1. The decision region $\Omega_i$ is shaded. $f(|\mathbf{x}|^2)$ has a sign as indicated by "+" and "-".

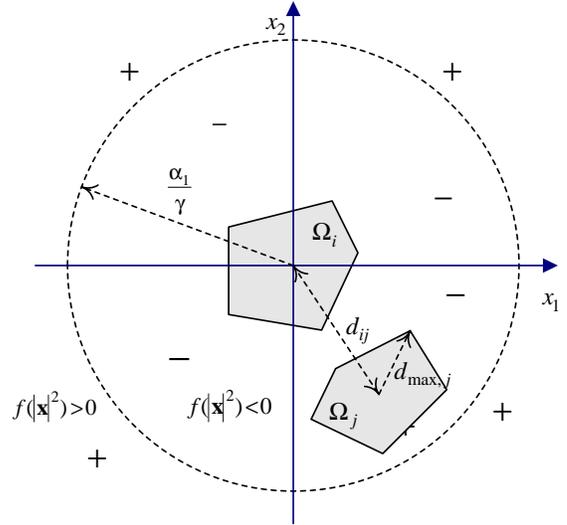

**Fig. 2.** Two-dimentional illustration of the problem geometry for Case 2.